# Actuated Reflector-Based Three-dimensional Ultrasound Imaging with Adaptive-Delay Synthetic Aperture Focusing

Yichuan Tang, Ryosuke Tsumura, *Member, IEEE*, Jakub T. Kaminski, Haichong K. Zhang, *Member, IEEE*

***Abstract*— Three-dimensional (3D) ultrasound (US) imaging addresses the limitation in field-of-view (FOV) in conventional two-dimensional (2D) US imaging by providing 3D viewing of the anatomy. 3D US imaging has been extensively adapted for diagnosis and image-guided surgical intervention. However, conventional approaches to implement 3D US imaging require either expensive and sophisticated 2D array transducers, or external actuation mechanisms to move a one-dimensional array mechanically. Here, we propose a 3D US imaging mechanism using actuated acoustic reflector instead of the sensor elements for volume acquisition with significantly extended 3D FOV, which can be implemented with simple hardware and compact size. To improve image quality on the elevation plane, we introduce an adaptive-delay synthetic aperture focusing (AD-SAF) method for elevation beamforming. We first evaluated the proposed imaging mechanism and AD-SAF with simulated point targets and cysts targets. Results of point targets suggested improved image quality on the elevation plane, and results of cysts targets demonstrated a potential to improve 3D visualization of human anatomy. We built a prototype imaging system that has a 3D FOV of 38 mm (lateral) by 38 mm (elevation) by 50 mm (axial) and collected data in imaging experiments with phantoms. Experimental data showed consistency with simulation results. The AD-SAF method enhanced quantifying the cyst volume size in the breast mimicking phantom compared to without elevation beamforming. These results suggested that the proposed 3D US imaging mechanism could potentially be applied in clinical scenarios.**

## I. INTRODUCTION

Due to its advantages in terms of safety, cost, portability, and high frame rate, ultrasound (US) imaging has been widely applied in disease diagnosis and image-guided surgical interventions. Compared to conventional two-dimensional (2D) US imaging, three-dimensional (3D) US imaging provides a volumetric field-of-view (FOV) and thus facilitates image evaluation and interpretation, as well as intervention planning with US images [1-3]. By far, 3D US imaging has been applied in fields such as cardiac imaging, prostate imaging, and fetus imaging during pregnancy [2]. Existing options for implementing 3D US imaging include using a 2D transducer array [1] (Fig. 1(a)), mechanically actuated one-dimensional (1D) array which conducts linear, tilting, or rotational scanning [2] (Figs. 1(b)-(c)), and sensor-tracking-free-hand scanning. Although the 2D array is capable of real-time 3D imaging, it requires expansive and sophisticated hardware, making it less accessible than a mechanically actuated 1D array. Among three mechanical scanning approaches of a 1D array, tilting and rotational scanning are useful in applications like prostate cancer imaging, where a side-firing linear array transrectal US probe is tilted about an axis parallel to the transducer face to collect a fan-like volume of images [4]. However, the image quality in tilting and rotational scanning deteriorates as the distance increases with respect to the tilting or rotational axis [3]. And especially for the rotational scanning, the image quality is sensitive to operator or patient motion since acquired 2D images intersect along the axis of rotation in the center of the 3D image, and motion will cause the axial pixels to have inconsistent values in different 2D images [3]. Linear scanning, which has a relatively more homogeneous resolution, is more suitable in applications where images need to be acquired across the patient's skin, such as breast, cardiac, and tumor vascular imaging [5]. However, a linear scanning 1D array usually requires an external actuation mechanism to hold the probe and track its position, e.g., a robotic arm [6]. While using

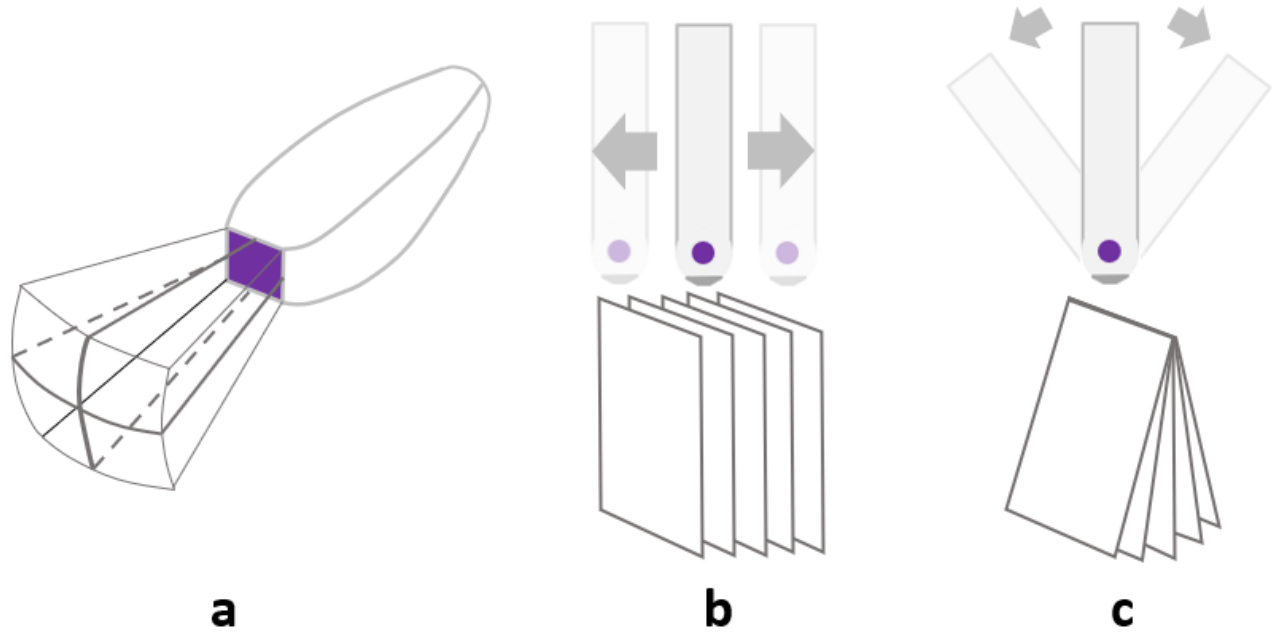

Fig. 1. **Conventional approaches to implement 3D US imaging.** (a) 2D array transducer. (b) 1D array with linear scanning. (c) 1D array with tilting scanning.

Manuscript received xxxx; revised xxxx; accepted xxxx. Date of publication xxxx; date of current version xxxx. This work was supported in part by Worcester Polytechnic Institute internal fund; in part by the National Institute of Health under Grant DP5OD028162, RO1CA134675. *(Corresponding author: Haichong K. Zhang.)*

Y. Tang, R. Tsumura, and J. T. Kaminski are with the Department of Robotics Engineering, Worcester Polytechnic Institute, Worcester, MA 01609 USA (e-mail: ytang7@wpi.edu; rtsumura@wpi.edu; jkaminski@wpi.edu).

H. K. Zhang is with the Department of Biomedical Engineering, Computer Science, and Robotics Engineering, Worcester Polytechnic Institute, Worcester, MA 01609 USA (e-mail: hzhang10@wpi.edu).

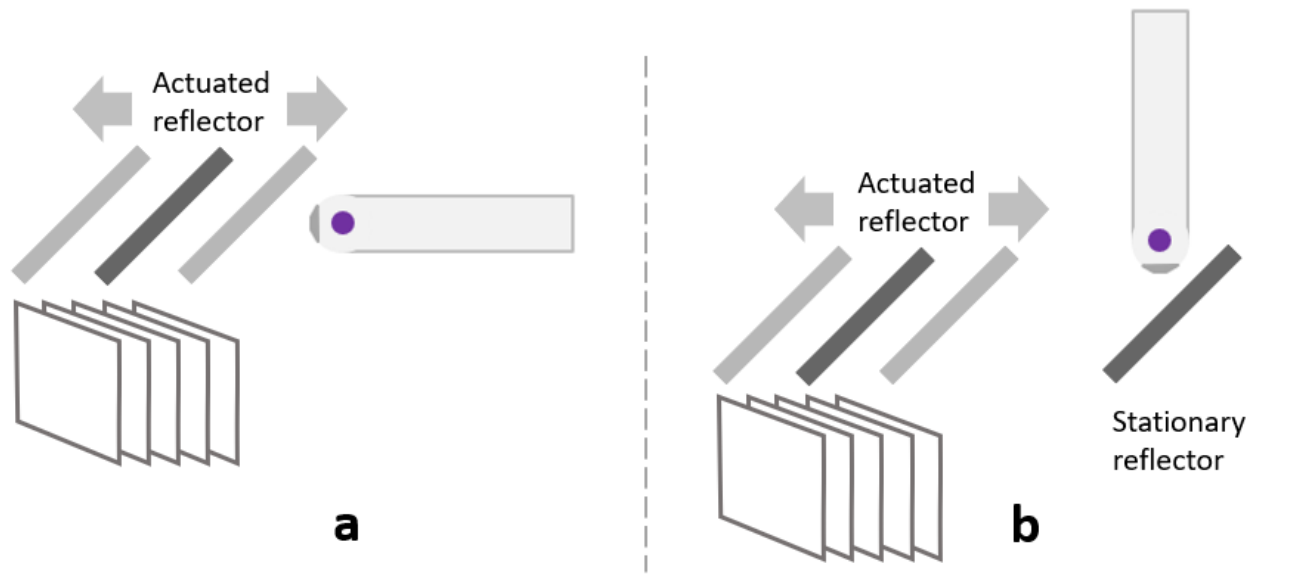

Fig. 2. **Proposed approaches to implement 3D US imaging using actuated acoustic reflectors.** (a) 1D array with a single actuated acoustic reflector for linear scanning. (b) 1D array with two acoustic reflectors, one stationary and one actuated for linear scanning.

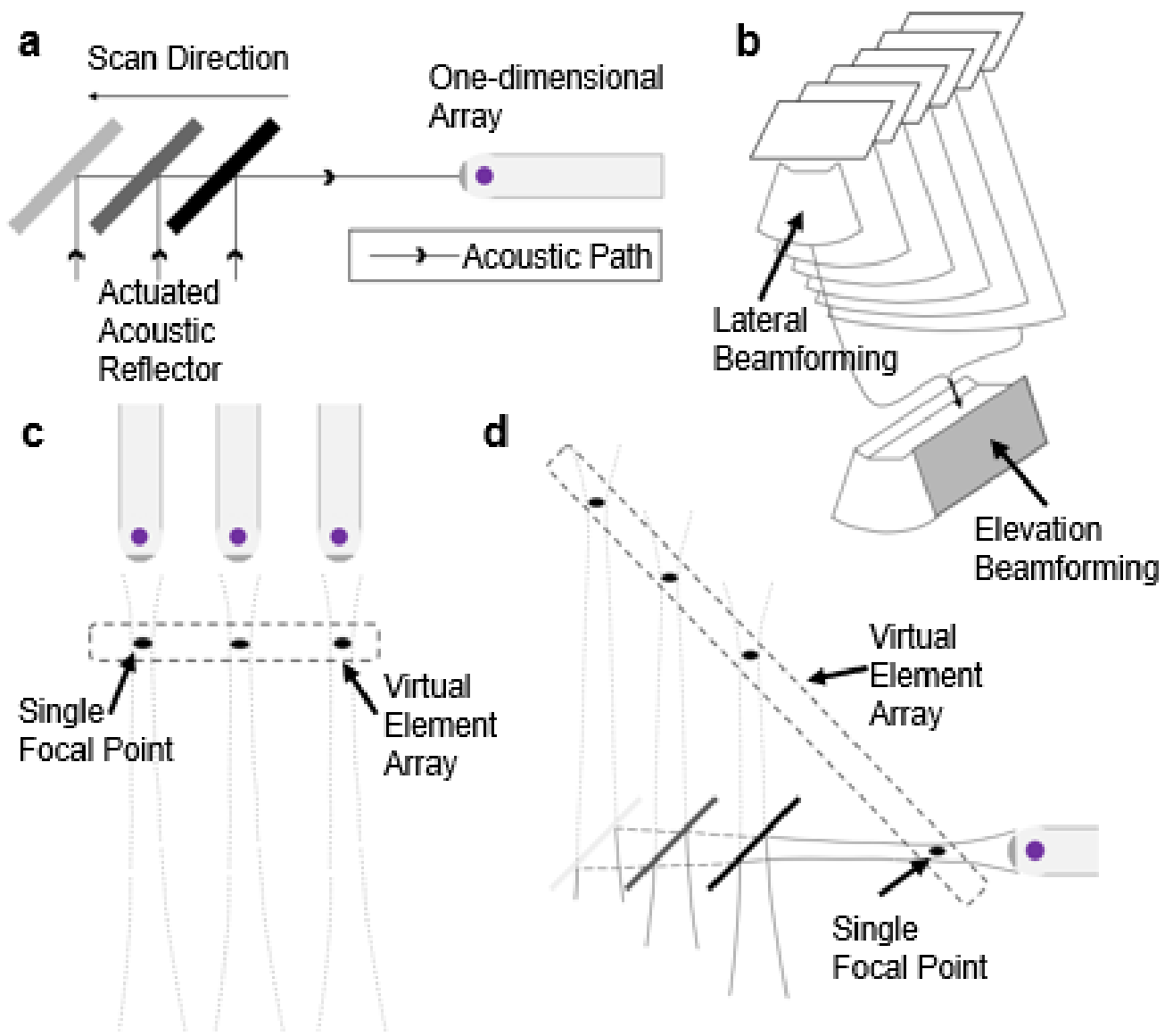

Fig. 3. **3D US image data acquisition and AD-SAF method.** (a) Acoustic path with actuated reflector. (b) Two-steps beamforming. (c) Virtual element array in conventional linear scanning 1D array. (d) Virtual element array in proposed imaging mechanism.

an external actuation unit provides extensive FOV and accurate tracking data, actuating the 1D array probe is not trivial due to its mass and geometry. Building a dedicated probe actuation unit can be costly and increase the system's overall size. A new linear scanning solution is needed to keep the advantages of the externally actuated 1D probe, i.e., homogeneous pixel size, large FOV, adaptability to commercially available clinical probes, and at the same time be low-cost and compact.

Acoustic reflectors were first introduced to change the orientation of acoustic waves in photoacoustic (PA) imaging devices [7-10], and previous work demonstrated their capability in facilitating fast 3D US imaging by rotating the acoustic beam to cover a fan-like 3D region [11]. In the context of linear scanning, volume data can be collected by actuating the reflector only, and no mechanism is needed to move the probe. The advantage of actuating the reflector lies in the reflector's smaller size and mass compared to the probe. The size of the reflector can be minimized to only cover the probe FOV, which indicates that an actuator with smaller size and power can be used to make the whole imaging system more compact. We here propose a new US imaging mechanism combining a 1D array and motor-actuated acoustic reflector to achieve 3D volume acquisition in a linear scanning manner (Fig. 2(a)). A commercially available clinical 1D array is placed horizontally and faces a glass-based acoustic reflector at a 45 ° angle with respect to the axial axis of the probe. The acoustic wave is reflected by 90° and thus travels in the vertical direction. 2D slices of radio frequency (RF) data are collected as the acoustic reflector moves along the elevation direction. A variation of the proposed image mechanism includes two acoustic reflectors, with one stationary reflector and one actuated reflector (Fig. 2(b)), where the 1D array is placed vertically. An advantage of the proposed imaging mechanism over the custom 3D imaging probe is the adaptability with most of the commercially available 1D arrays. Since the actuator lies inside the imaging mechanism, size is reduced significantly and can suit hand-held operations.

In addition, RF data are beamformed on the elevation plane to improve the quality of the 3D US image. The acoustic lens focal point is regarded as the virtual element to synthesize the elevation aperture information, and delay-and-sum (DAS) is applied to focus the signals. Yet, the conventional synthetic aperture focusing (SAF) is not directly applicable because the proposed configuration varies the effective focal point as the distance between the transducer array and reflector changes, i.e., scatters at the same physical depth have different axial positions in collected 2D RF data. Therefore, to accommodate the unique acoustic path geometry in the proposed imaging mechanism, we introduce an adaptive-delay synthetic aperture focusing (AD-SAF) method, which extends the conventional SAF by adding variable depth function in the formulation.

In this paper, we present the theory of acoustic wave reflection and the principle behind AD-SAF in Section II. In Section III, we discuss the validation of AD-SAF through simulation and prototype imaging system implementation, as well as the setup of the prototype in experiments. These validation results are presented in Section IV, which is followed by a discussion in Section V and conclusions in Section VI.

## II. Theories and Methods

This section introduces the theory of acoustic wave reflection in the proposed imaging mechanism, the 3D US data acquisition pipeline, and the AD-SAF method aiming to improve image quality on the elevation plane.

### *A. Reflection of Acoustic Wave*

Acoustic reflectors have been proven to be able to reorient acoustic waves with minimal energy loss. Previous work [11-13] demonstrated successful acoustic reflector application in the 3D US imaging. A quantified study investigating the amount of distortion in US beam caused by acoustic reflector was carried out by Dong et al. [11]. Acoustic reflectors are usually made of materials such as glass and metal, whose acoustic impedance are significantly different from that of medium of acoustic wave propagation, like water and gel, thus create a high acoustic reflection coefficient on the media interface. The reflection of acoustic waves in the proposed imaging mechanism is illustrated in Fig. 3(a). Here, we examine if there is a loss of acoustic energy caused by the reflector in the proposed mechanism based on Snell's law [14]. The acoustic incidence angle $\theta_i$ is related to the refraction angle $\theta_r$ as described in (1):

$$\frac{\sin(\theta_i)}{\sin(\theta_r)} = \frac{c_1}{c_2} \tag{1}$$

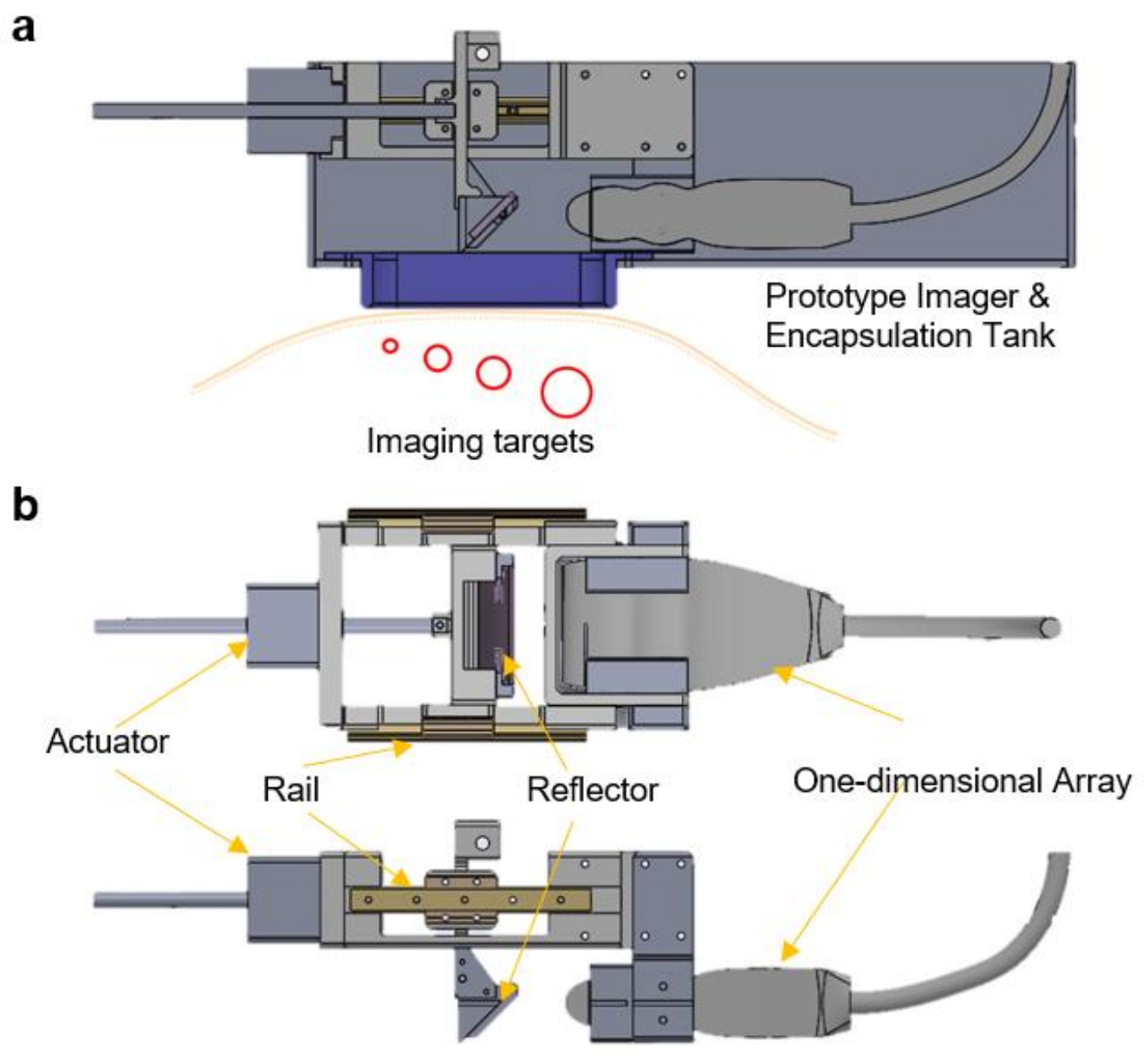

Fig. 4. **Prototype of 3D US imager based on a single actuated acoustic reflector.** (a) Concept of prototype imager inside encapsulation water tank scanning ultrasound imaging targets. (b) Design details of the prototype imager.

The acoustic wave propagates from Medium 1 (speed of sound is $c_1$) into Medium 2 (speed of sound is $c_2$). Although only longitudinal wave approaches the media interface, there can be both longitudinal and shear waves transmitting into Medium 2. Speeds of longitudinal and shear waves in Medium 2 are denoted as $c_{2d}$ and $c_{2s}$, respectively. The first critical condition occurs when the refraction angle of the longitudinal wave becomes 90°, and the corresponding incident angle is recognized as the first critical angle $\theta_{c1}$, as in (2). If $\theta_i > \theta_{c1}$, no longitudinal wave transmits through the interface.

$$\theta_{c1} = \text{asin}(\frac{c_1}{c_{2d}}) \tag{2}$$

The second critical condition occurs when the refraction angle of shear wave becomes 90°, and the second critical angle $\theta_{c2}$ is defined as,

$$\theta_{c2} = \text{asin}(\frac{c_1}{c_{2s}}) \tag{3}$$

If $\theta_i > \theta_{c2}$, all acoustic energy arrives on the media interface will be reflected. In the proposed mechanism, the acoustic reflector is made of glass and immersed in water. Given these conditions, $c_1$, $c_{2d}$, $c_{2s}$ are assumed to be 1480 $m/s$, 5100 $m/s$, 2840 $m/s$, respectively [15]. It could be easily computed using (2) and (3) that $\theta_{c1} \cong 16.87\,°$ and $\theta_{c2} \cong 31.41\,°$. Since the incident angle is 45 ° in the proposed imaging mechanism, and is larger than both critical angles, there will be no loss in acoustic energy in acoustic reflection theoretically.

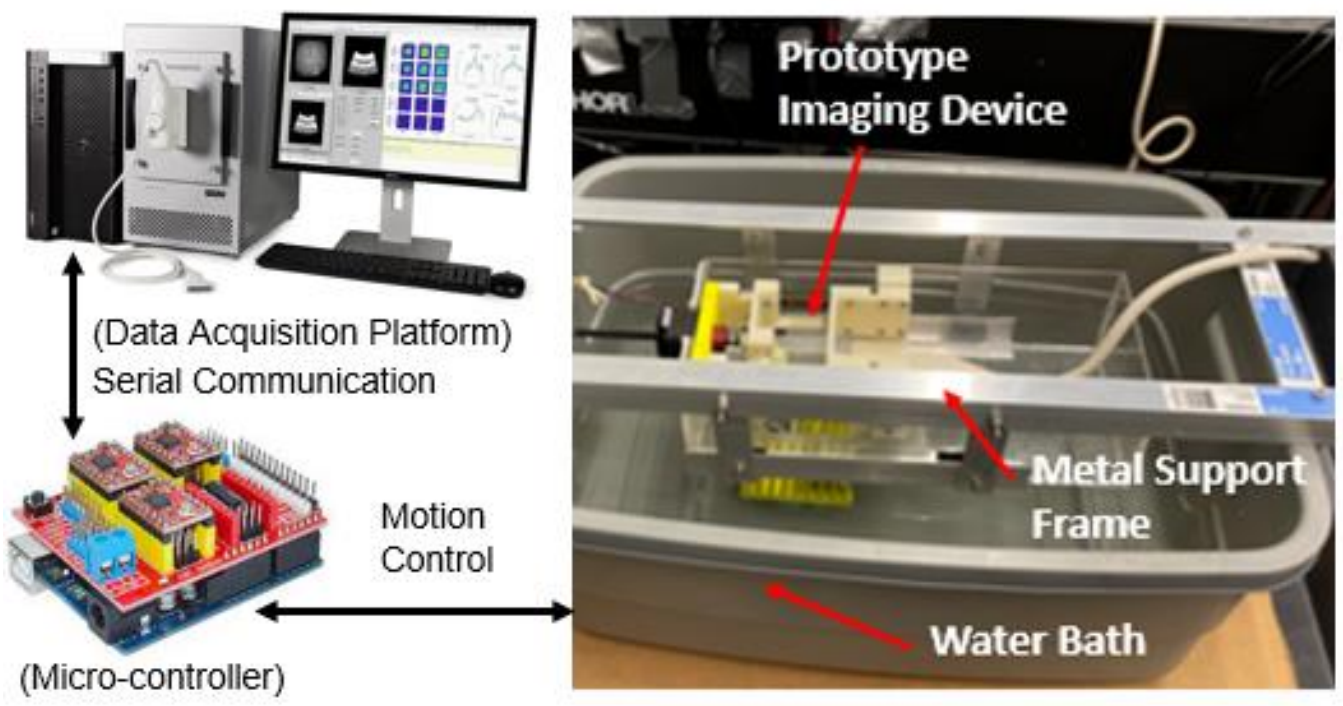

Fig. 5. **Prototype 3D US imaging system.** Communication between data acquisition and motion control is implemented via serial portal. The imaging device is fixed using a metal frame and is partially submerged in a water bath.

### *B. 3D US Data Acquisition*

The actuated reflector scans with fixed step size, and thus location information of each acquired RF slice is available. After beamforming on the lateral plane, RF slices are stacked for beamforming on the elevation plane, and such a two-steps beamforming is illustrated in Fig. 3(b). SAF used in conventional actuated 1D array 3D US imaging system is extended to adapt to the geometry of the variable acoustic path, resulting in the proposed AD-SAF method proposed to focus signal on the elevation plane.

### *C. Adaptive Delay Synthetic Aperture Focusing (AD-SAF)*

To improve image quality on the elevation plane, we propose the AD-SAF method. Nikolov and Jensen [16] first applied synthetic aperture focusing (SAF) based on the concept of virtual element for elevation beamforming in a 3D US imaging system, which is implemented with the linearly translated 1D array. In the proposed imaging mechanism, distance from the probe to the reflector is variable, resulting in a changing axial position in received signals for scatters with the same physical depth. Hence, a extended version of SAF method incorporating the varying focal point is needed.

In the conventional SAF method, the trajectory of a single elevation focal point forms a virtual element array parallel to the horizontal direction, as shown in Fig. 3(c). Fig. 3(d) shows the virtual element array of the proposed imaging mechanism: on each acquisition position, the virtual element is found using the mirror-reflection symmetry. Based on the geometry of the virtual element array, the AD-SAF method is formulated. The steps of AD-SAF are summarized in (4)-(9). Since only signals from the space below the reflector include echoes from the imaging subject, the origin of RF data received at step $n$, or $r_n$, is defined as in (4),

$$r'_n(t) = r_n\left(t + \frac{2 * \left(d_{tr,m} + n\Delta\delta\right)}{c}\right) \quad (4)$$

where $d_{tr,m}$ represents the distance from the surface of the US transducer (denoted as $tr$) to the acoustic reflector (denoted as $m$) when the reflector is in the home position, $t$ is the time, $c$ is the speed of sound, and $\Delta\delta$ is the step size of the linear motion. After defining the origin, RF data are focused on the elevation plane using DAS. The relationship between with and without beamforming RF data is formulated in (5).

$$r_{bf}(i,j) = \sum_{n=1}^{N_p} r'_n\left(\frac{2 * d_{ijf}(i_n, i, j, n)}{c}\right) \quad (5)$$

where $r_{bf}$ is the beamformed RF data on the elevation plane, $i$ and $j$ are pixel indices in the axial direction and the elevation direction. The distance from the location of pixel $(i,j)$ to the elevation focal point of the linear array is denoted as $d_{ijf}$, which can be computed with (6)-(9).

$$d_{ijf}(i_n, i, j, n) = dist(i_n, i, j, n) - d_{comp}(n) \quad (6)$$

$$dist(i_n, i, j, n) = \sqrt{\left((i - i_n) * \Delta\delta\right)^2 + (j * ss + d_{comp}(n))^2} \quad (7)$$

$$d_{comp}(n) = d_{tr,m} + n * \Delta\delta - f \quad (8)$$

$$ss = \frac{c}{2f_s} \quad (9)$$

In (9), $ss$ is the axial sampling spacing, $f_s$ is the sampling frequency, and $f$ is the elevation focal depth. Since the origin of RF signals is defined in (2), $d_{ijf}$ needs to be compensated by $d_{comp}$, as defined in (8), to represent physical depth.

## III. Implementation

This section provides implementation details of simulation studies and setup for imaging experiments conducted to validate the AD-SAF method.

### A. Simulation Setup

Simulation studies are implemented in MATLAB with Field II simulation platform [17] to validate the AD-SAF method. A model of ATL L12-5 38 mm linear array, which is the probe installed in the prototype, is set up in the simulation. The central frequency is configured to be 5 MHz, and the sampling frequency is set as 40 MHz. The linear motion step size is set as 200 µm with a scanning range of 38 mm. The elevation and axial location of imaging targets are adjusted to simulate the varying length of the acoustic path on each acquisition position. In addition, 10 dB white noise is added to RF data. Point targets are used to evaluate the point spread function (PSF) of the 3D US imaging system, and a phantom of three cysts (each cyst has a diameter of 10 mm) is used to investigate the potential in imaging human anatomy. Key parameters for the simulation study setup are summarized in Table 1.

### B. Experiment Setup

The prototype 3D US imager and imaging system are shown in Fig. 4 and Fig. 5, where the data acquisition platform

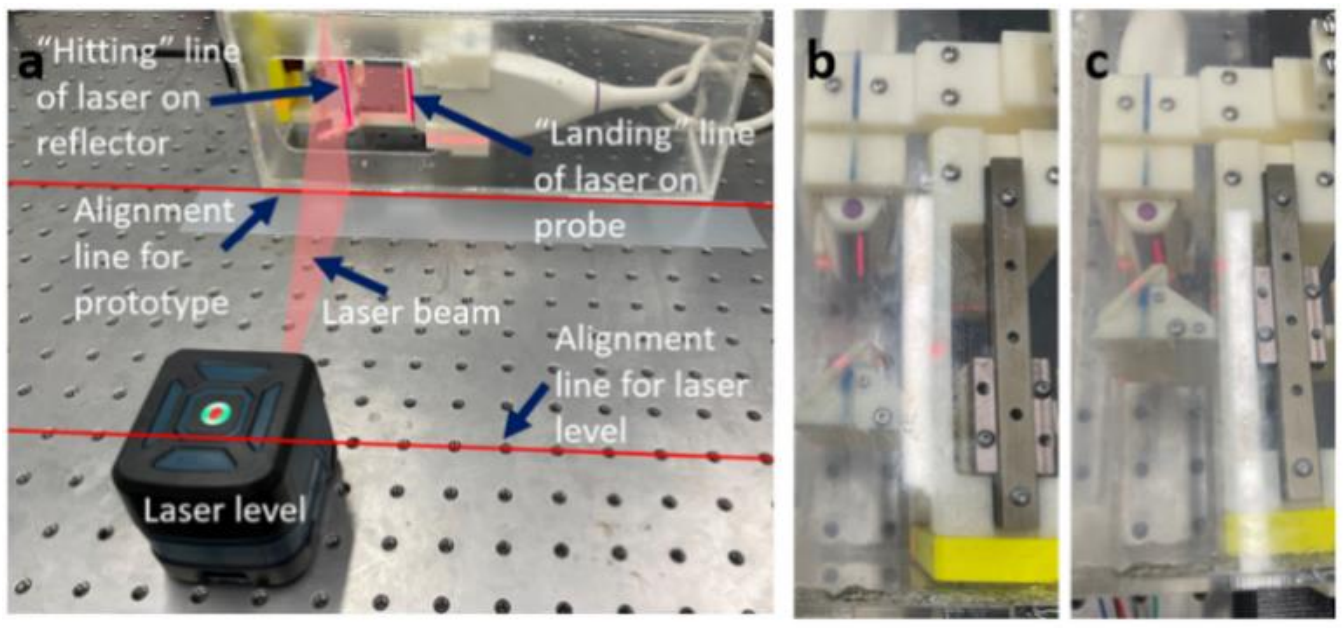

Fig. 6. **Calibration of acoustic reflector angle.** (a) Reflector angle calibration setup with a laser level. (b) Landing line of laser on the probe with the reflector on a far-end position with respect to the probe. (c) Landing line of laser on the probe with the reflector on a near-end position.

(Vantage 128, Verasonics Inc., WA, USA) sends motion command to the actuator (Non-captive NEMA 14, PCB Linear, IL, USA) via serial port after data acquisition on each sampling position. The linear motion step size is 200 µm, and the maximum elevation scan is 38 mm. The linear array is configured to emit plane-wave [18] at 7 different angles evenly spaced in the range of -18° and +18°. RF data are beamformed on the lateral plane using the embedded program on the data acquisition platform. Two types of phantoms are scanned: fishing-wire phantom (200 µm diameter) and breast elastography phantom (Part Number 1552-01, CIRS), aiming to evaluate the PSF and human tissue imaging potential of the prototype. The prototype imager is supported by a metal frame and fixed on top of a water bath. Phantoms and the part of the prototype are submerged in the water tank, including a 1D array and actuated acoustic reflector.

TABLE 1
Simulation Study Parameters

| Parameters | Notation | Value | Unit |
|---|---|---|---|
| Speed of sound | $c$ | 1480 | $m/s$ |
| Central freq. | $f_0$ | 5 | $MHz$ |
| Sampling freq. | $f_s$ | 40 | $MHz$ |
| Num. elements | $N_{xdc}$ | 192 | - |
| Element pitch | $p$ | 197.9 | $\mu m$ |
| Element kerf | $kerf$ | 25 | $\mu m$ |
| Element height | $h$ | 4.5 | $mm$ |
| Elevation focus | $f$ | 20 | $mm$ |
| Scan range | $r$ | 38 | $mm$ |
| Scan step | $\Delta\delta$ | 200 | $\mu m$ |

### C. Calibration with Optical Method

The ideal angle of the acoustic reflector with respect to the acoustic path is 45 °; however, the actual angle might have an error due to 3D printing and installation. Calibration is carried out to validate if the angle matches 45 °. A laser level is used to visualize the acoustic path by orienting the laser beam towards the reflector, and the reflected beam lands on the acoustic lens of the US probe, as shown in Fig. 6 (a). The beam landing position on the transducer can be used to verify if an error exists in the reflector angle. If the beam landing location changes when the reflector moves, it means that an error exists in the reflector angle. It is found that the beam landing location is consistent when the reflector is placed on different elevation positions, as Figs. 6 (b)-(c) demonstrate, indicating that the error in the reflector angle is minimized.

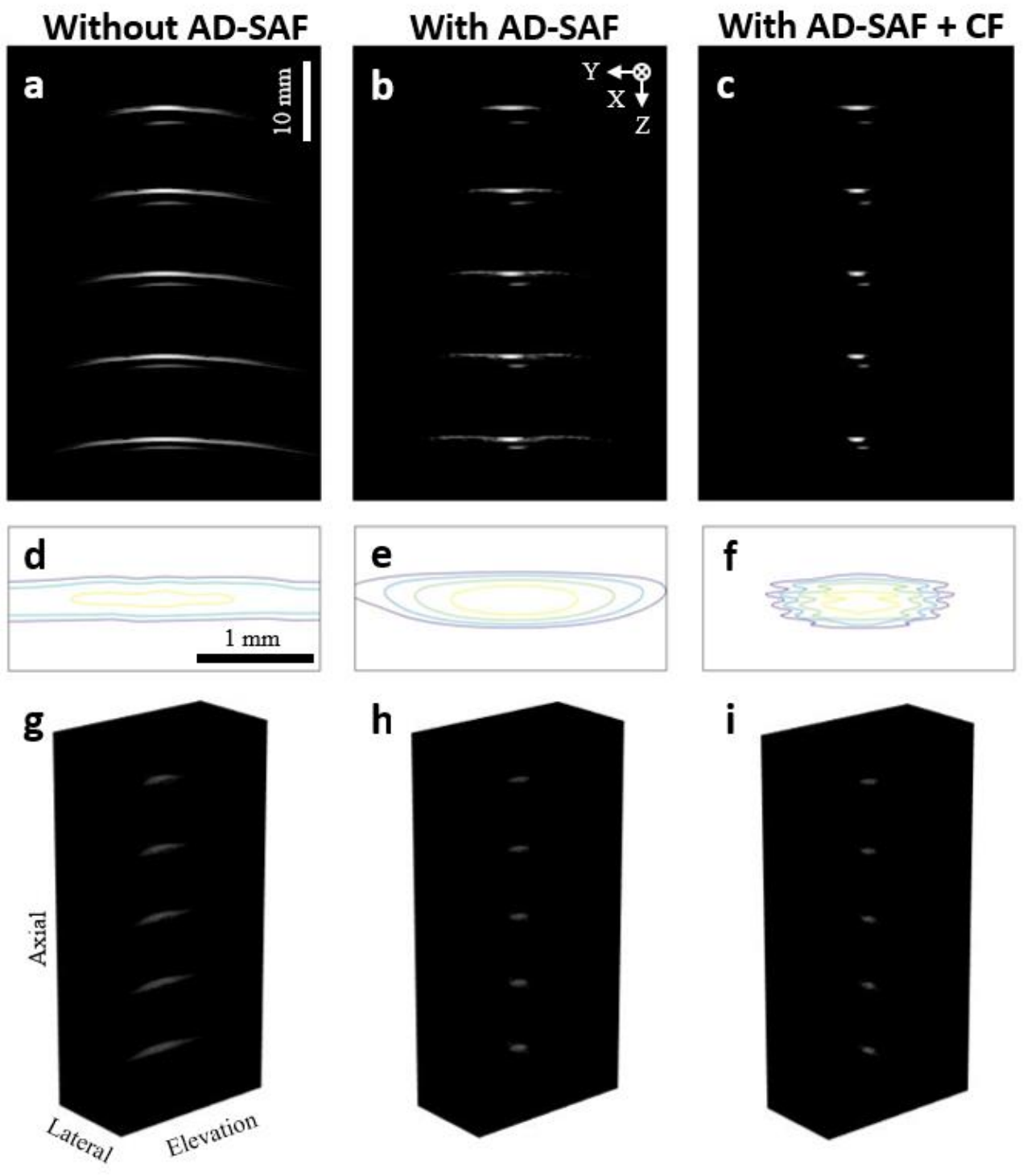

Fig. 7. **Simulated PSF of point targets on elevation plane, dynamic range 40 dB.** (a) Point targets without AD-SAF on the central lateral position. (b) Point targets with AD-SAF, using DAS. (c) Point targets with AD-SAF, with CF. (d) Contour plots of PSF without AD-SAF at 27mm depth, the innermost contour is at level of -3dB, and the interval between neighboring contours is 3dB. (e) contour plots of PSF with AD-SAF, using DAS. (f) contour plots of PSF with AD-SAF, with CF. (g) 3D volume of PSF without AD-SAF segmented at -20dB. (h) 3D volume of PSF with AD-SAF, using DAS. (i) 3D volume of PSF with AD-SAF, with CF.

## IV. Results

Simulated data are used to evaluate the effects of AD-SAF on improving image quality on the elevation plane and to study the effects of system parameters such as motion step size and central frequency. Data from step size and central frequency studies are referred to when configuring the prototype. Experimental data show consistency with simulation results in both cases of point targets and the breast elastography phantom. Apart from conventional delay-and-sum, coherent factor (CF) [19-20] was implemented to suppress side lobes.

### *A. Simulated Points and Cysts*

In the simulation with point targets, elevation images in the central lateral position with and without AD-SAF are displayed in Figs. 7(a)-(c) with 40 dB dynamic range. The point target at 27 mm depth is chosen for further examination, whose contour plots are shown in Figs. 7(d)-(f). The quality of elevation images is evaluated with the full width at half maximum (FWHM) and the signal-to-noise ratio (SNR). PSF in 3D space is segmented at -20dB and visualized in Figs. 7(g)-(i) to illustrate the improvement in 3D image resolution. In the simulation with cysts phantom, images on the central lateral position are displayed in Fig. 8 with 60 dB dynamic range to demonstrate the potential of AD-SAF in improving the image quality of human anatomy. Contrast-to-noise ratio (CNR) [21-22] is computed to evaluate the change of image quality after applying AD-SAF. The formula of CNR is given in (10), where $\mu_1$ and $\mu_2$ are mean pixel values of ROI and background region, and ${\sigma_1}^2$ and ${\sigma_2}^2$ are corresponding standard deviation values.

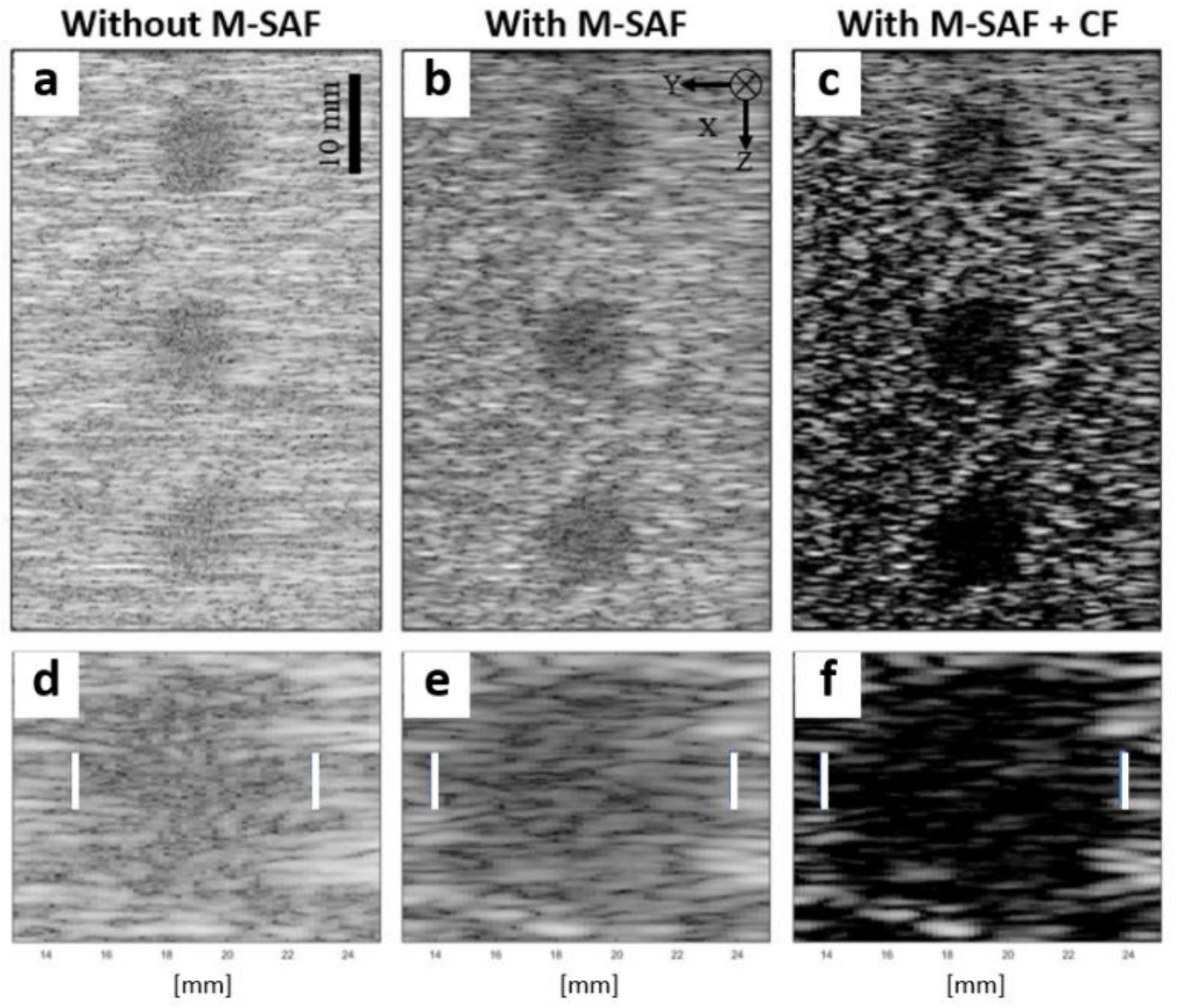

Fig. 8. **Simulated cysts phantom on elevation plane, dynamic range: 60 dB.** (a) Cysts without AD-SAF. (b) Cysts with AD-SAF, delay-and-sum. (c) Cysts with AD-SAF, with CF. (d) Enlarged cyst on the middle position in (a). (e) Enlarged cyst on the middle position in (b). (f) Enlarged cyst on the middle position in (c).

$$CNR = \frac{|\mu_1 - \mu_2|}{\sqrt{{\sigma_1}^2 + {\sigma_2}^2}} \tag{10}$$

TABLE 2
The quantified results of points and cysts simulations

<table>
<tr><td colspan="5">FWHM and SNR of simulated PSF at 27 mm depth</td></tr>
<tr><td></td><td>Without AD-SAF</td><td>With AD-SAF, DAS</td><td colspan="2">With AD-SAF, DAS+CF</td></tr>
<tr><td>FWHM (mm)</td><td>4.98</td><td>2.60</td><td colspan="2">1.40</td></tr>
<tr><td>SNR (dB)</td><td>42.00</td><td>48.38</td><td colspan="2">49.34</td></tr>
<tr><td colspan="5">CNR and measured diameter of simulated cyst at 30 mm depth</td></tr>
<tr><td colspan="2">CNR (*10^-2)</td><td>0.57</td><td>0.71</td><td>0.67</td></tr>
<tr><td colspan="2">Diameter (mm)</td><td>8.00</td><td>10.00</td><td>10.00</td></tr>
</table>

### *B. System Optimization and Tolerance Evaluation*

Point scatters simulation is used as a tool for imaging parameter optimization [23]. In this paper, the central frequency $f_0$ and the elevation motion step size $\Delta\sigma$ are varied to investigate their effects on the image quality. In the central frequency study, the elevation motion step size is kept as 200 µm, and the central frequency varies from 1 MHz to 9 MHz; while in the motion step size study, the central frequency is fixed as 5 MHz, and the elevation motion step size is changed from 200µm to 1800 µm. FWHM and SNR were used as criteria to evaluate the image quality from each combination of the two parameters. In the case of step size study, images are interpolated in the elevation direction for the convenience of quantification. The central frequency study (Figs. 9(a), (c)) suggests using a higher central frequency could benefit both the resolution and SNR. However, using a central frequencies higher than 5 MHz does not bring significant improvements. Also, the trend in the step size study (Figs. 9(b), (d)) suggests an upper limit for step size around 1000 µm, above which the spatial sampling along the elevation direction is insufficient. Such an indication corresponds to the upper limit of the elevation sampling step for synthetic aperture system derived from the Nyquist sampling principle [24], as shown in (11). Since the element height $h$ is set to be 4.5 mm, the motion step size should be smaller than 1125 µm (1/4 of 4.5mm), and according to the step size study results, a small step size brings better resolution after applying AD-SAF. As a result, the central frequency in the prototype is chosen as 5 MHz, and the elevation step size is chosen as 200 µm.

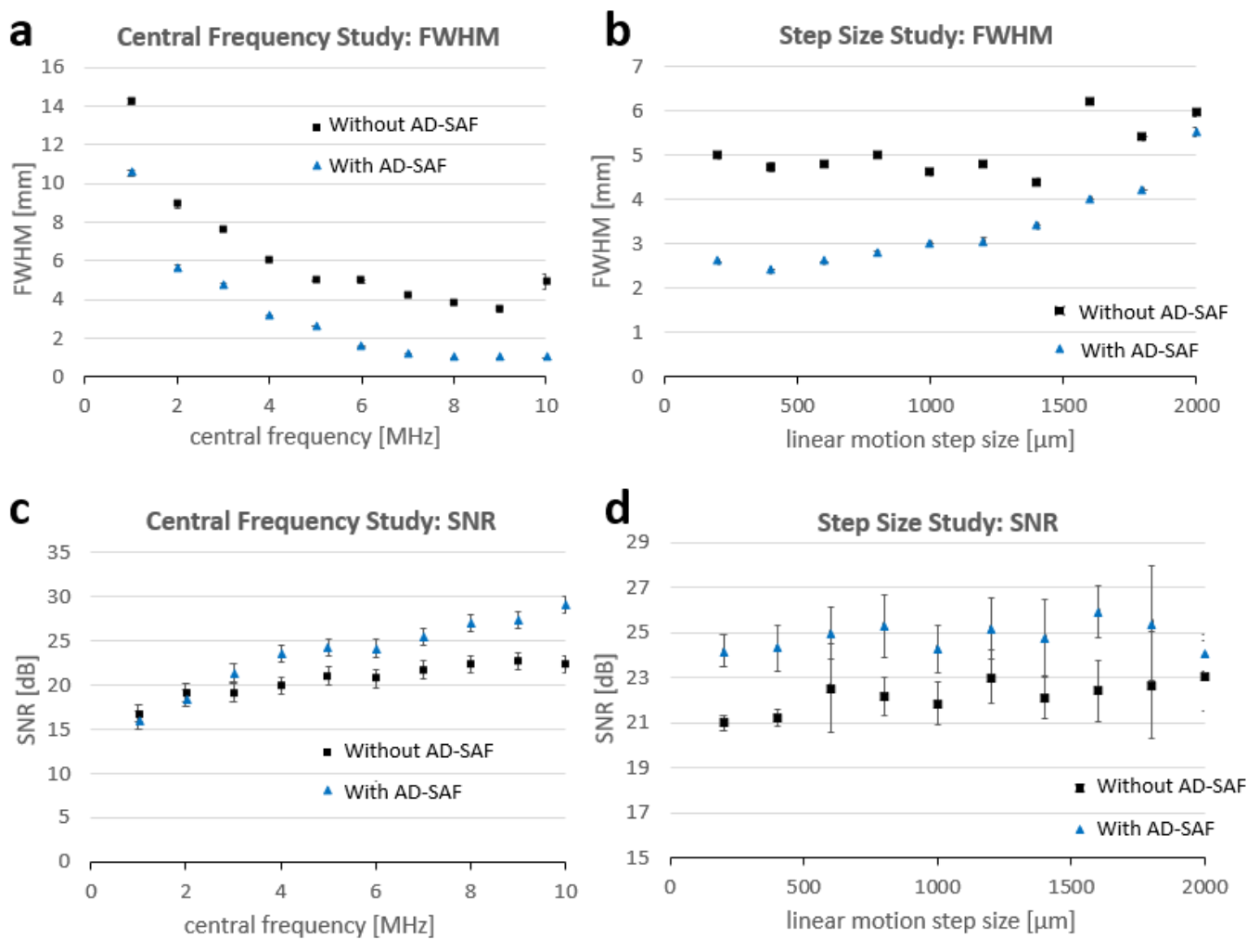

Fig. 9. **Central frequency and motion step size optimization for AD-SAF.** (a) FWHM of different central frequencies without and with AD-SAF, DAS. (b) FWHM of different elevation step sizes. (c) SNR of different central frequencies. (d) SNR of different elevation step sizes.

$$\Delta\delta \leq \frac{h}{4} \qquad (11)$$

## *C. Wires Phantom*

In the wire phantom imaging experiment, a phantom of four fishing wires (100% Nylon, 200 µm diameter) is scanned with fishing wires aligned with the lateral direction of the probe to simulate point targets on the elevation plane. The with and without AD-SAF images on the central lateral position are shown in Figs. 10(a), (c), (e), displayed with a 40 dB dynamic range. The visualized volume data of fishing wires segmented at -20 dB are presented in Figs. 10(b), (d), (f). The displayed FOV is 38 mm (lateral) by 38 mm (elevation) by 25 mm (axial). FWHM and SNR of four wire targets are computed and summarized in Table 3.

TABLE 3
The quantified image quality of wires phantom

| FWHM and SNR of four wires phantom on elevation plane | | | |
|---|---|---|---|
| Mean (Std.) | Without AD-SAF | With AD-SAF, DAS | With AD-SAF, DAS+CF |
| FWHM (mm) | 4.4 (0.35) | 2.13 (0.61) | 0.8 (0.20) |
| SNR (dB) | 38.69 (4.01) | 43.87 (3.70) | 51.07 (1.14) |

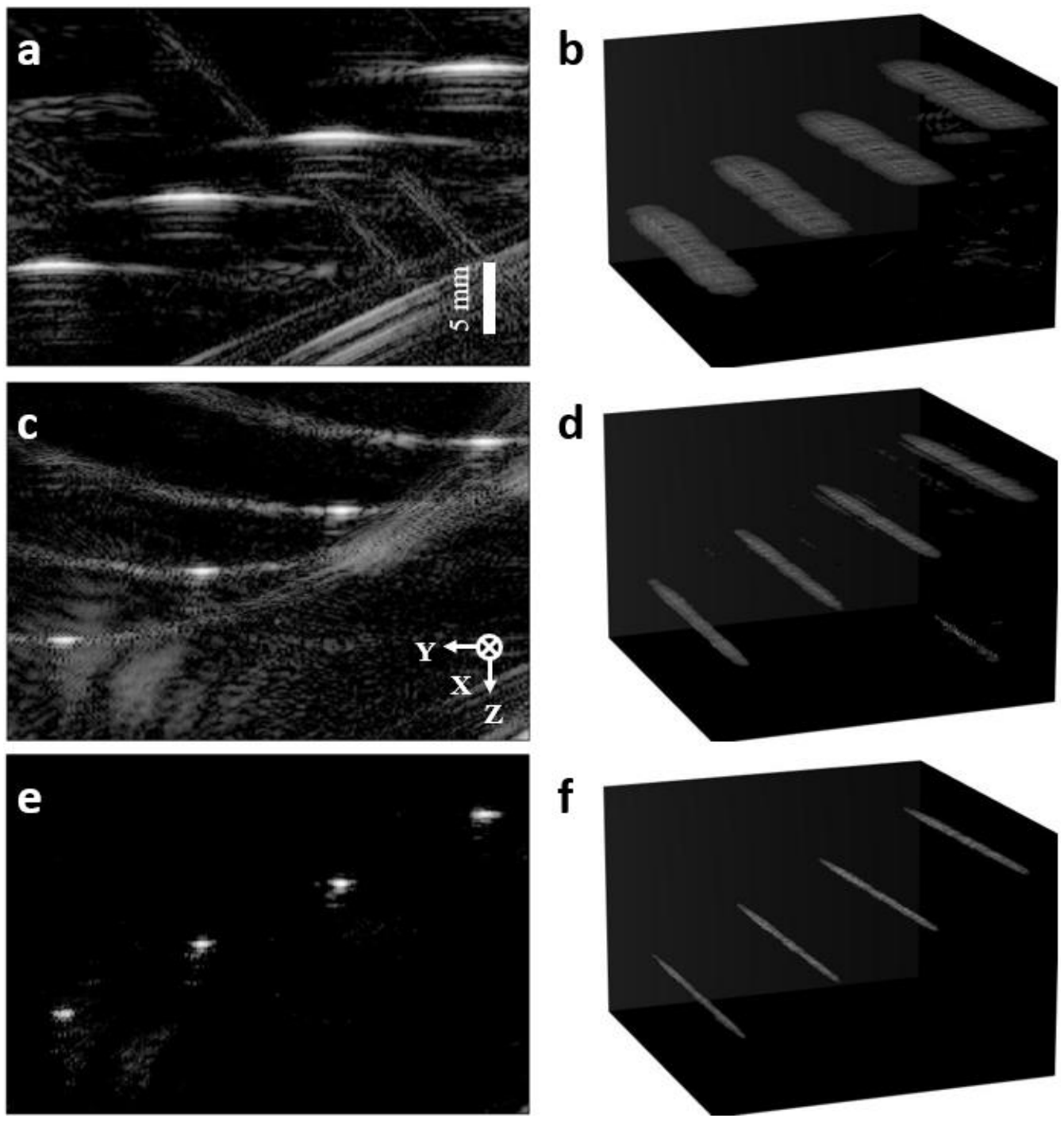

Fig. 10. **Experimental data of wires phantom, dynamic range: 40 dB.** (a) Image of wires phantom on the elevation plane without applying AD-SAF. (b) 3D visualization of fishing-wire phantom without AD-SAF. (c) Image of wires phantom with AD-SAF, DAS. (d) 3D visualization of fishing-wire phantom with AD-SAF, DAS. (e) Image of wires phantom with AD-SAF, DAS with CF. (f) 3D visualization of fishing-wire phantom with AD-SAF, DAS with CF

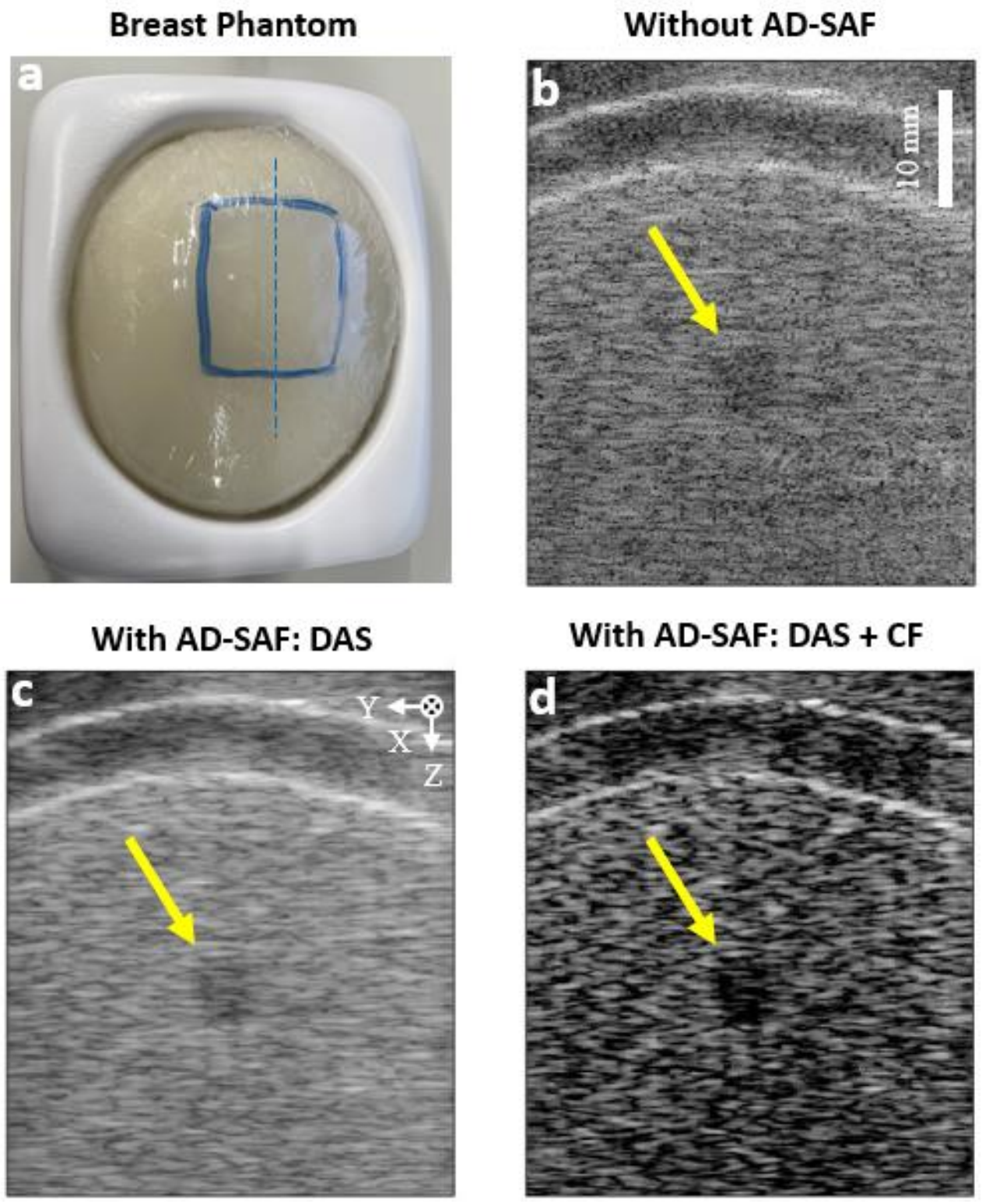

Fig. 11. **Experimental data of breast phantom, dynamic range: 70 dB.** (a) Photo of the breast elastography phantom, with rectangle indicating scanned region and dashed line indicating the position of displayed 2D images. (b) Image of breast phantom without AD-SAF. (c) Image of breast phantom with AD-SAF, DAS. (d) Image of breast phantom with AD-SAF, DAS with CF.

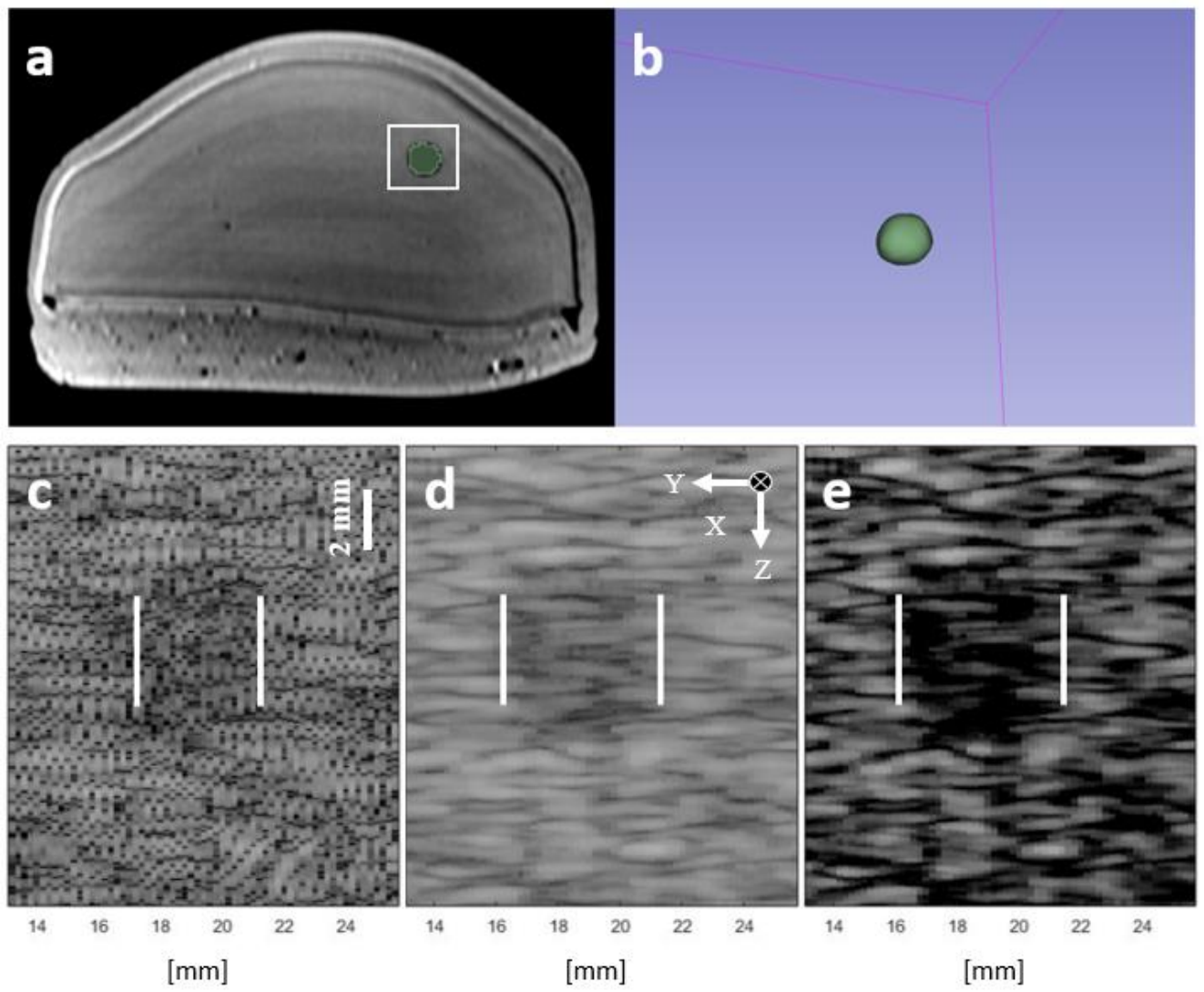

Fig. 12. **Evaluation of breast phantom results.** (a) MRI image of the cyst region in interest. (b) Segmented cyst in 3D view. (c)-(e) Zoom-in cyst region in Figs. 10 (b)-(d).

### *D. Breast Phantom*

A breast elastography phantom with cystic masses is imaged in this experiment. One of the cystic masses is chosen as the imaging target, and its neighboring region is scanned. Images on the elevation plane are visualized in Fig. 11 for both with and without AD-SAF cases with a 70 dB dynamic range. 3D US data of the scanned volume are visualized in Fig. 13, with the cyst located in the central location. MRI scanning of the breast phantom is conducted to evaluate the performance of AD-SAF on breast phantom data. The diameter of the cyst in the breast phantom is estimated according to segmented MRI cyst volume and is referred to as the ground truth to compare with the cyst diameter estimated from the US volume. Cyst region in MRI image is segmented using the level tracing function in 3D Slicer [25], and then the diameter of the cyst is computed based on the volume of the cyst (assumed to be a sphere), which turns out to be 7.14 mm. MRI image of the cyst in breast phantom and segmented cyst volume are demonstrated in Figs. 12 (a)-(b). The CNR between cyst region and background is computed for with and without AD-SAF US images, and the diameter of cyst is measured from US images (as shown in Figs. 12 (c)-(e)) and compared with ground truth value estimated from MRI image, as summarized in Table 4.

TABLE 4
The measured CNR and cyst diameter of imaged breast phantom

| Diameter of cyst on elevation plane, in the elevation direction | | | |
|---|---|---|---|
| | Without AD-SAF | With AD-SAF, DAS | With AD-SAF, DAS+CF |
| CNR (*10^-2) | 1.06 | 1.75 | 2.64 |
| Diameter (mm) | 4.20 | 5.20 | 5.20 |
| Percent error (%) (MRI ground truth: 7.14mm) | 41.18 | 27.17 | 27.17 |

## V. DISCUSSION

We validated the 3D US imaging mechanism proposed in this paper for its capability in obtaining volume data and improved image quality on the elevation plane, benefiting from the proposed AD-SAF method, in both simulation and experiments. We demonstrated that the proposed imaging mechanism could adapt to a clinically available probe and does not require dedicated probe design and fabrication like existing 3D US imaging systems. The range of the reflector motion can be adjusted to fit the requirement of different imaging applications. Since the mass of the reflector is smaller than a probe (the mass ratio between reflector and probe is a around 1:15 according to our measurement), a higher volume rate is potentially achievable compared to actuating the 1D array linearly. Such a reflector-based imaging mechanism could also be valuable in US image-guided needle intervention [29-31] by opening a hole on the reflector, allowing the needle to pass [32]. In this way, the needle path is registered to the image plane to provide stable visualization of the needle during the insertion process.

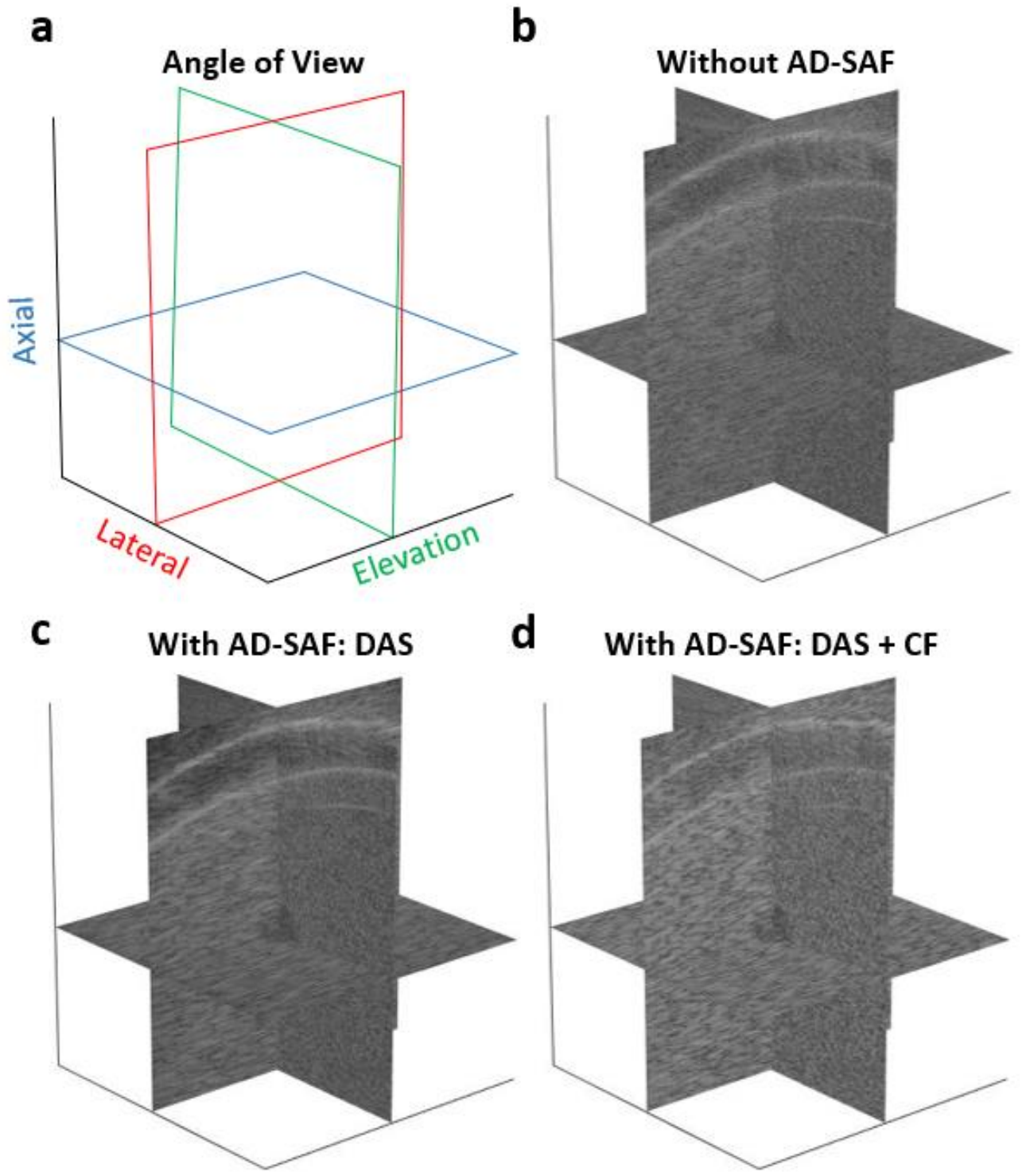

Fig. 13. **3D volume scanned from breast phantom, dynamic range 70 dB** (a) Angle of view of the 3D visualization. (b) 3D visualization of volume without AD-SAF. (c) 3D visualization with AD-SAF, DAS. (d) 3D visualization with AD-SAF, DAS with CF.

The concept of treating transducer focal point as virtual element in SAF processing was first introduced by Passman and Ermert [33] to overcome the limitation in lateral resolution in the traditional B-mode image [34]. With its application in conventional linearly translated 1D array to improve image quality on the elevation plane successfully demonstrated [16], SAF has been applied in various forms of 3D US imaging systems, including a rotationed phased array probe [35] and robot-tracked convex probe [6] to improve 3D data quality. This paper further extends the applicability of SAF in more flexible settings by adapting to the variable virtual focal point.

Some limitations should be addressed before clinical application. Due to the existence of the acoustic reflector, the US probe cannot get in direct contact with the patient. As a result, the imaging device must be emersed in the acoustic coupling medium and encapsulated. The interface layer (bottom) of the encapsulation can attenuate acoustic waves and strength of echos from the patient can be weakened. Multiple reflection artifacts will also be caused by the water-bottom and bottom-tissue acoustic interfaces. Previous work using acoustic reflectors for hand-held PA/US imaging also reported similar issues to be solved [9-11]. Potential encapsulation of the proposed imaging mechanism can be built referring to encapsulations used in mechanically swept 3D US probe [3] or automated breast ultrasound imaging system (Invenia ABUS 2.0, General Electric Healthcare) [26]. Materials with similar acoustic impedance to water can be used for the interface layer [27] to minimize multiple reflection artifacts, such that the reflection coefficient of acoustic waves can be reduced. Also, advanced signal processing methods are helpful in removing multiple reflection artifacts in the post-processing phase [28]. Such an encapsulation solution can be generalized and will benefit other imaging mechanisms involving acoustic reflectors. Apart from issues related to encapsulation, the current scanning time of the prototype is around 1 minute, which limits the 3D imaging volume rate. Further research will be carried out to improve the volume rate of the proposed imaging system, e.g., using a linear mechanism with higher motion speed, such as piezo linear actuators, and enabling continuous scanning without stopping on each sampling position along the elevation direction. Also, If the speed of reflector motion is high, blurring could occur and affect the quality of data acquired. A potential topic to investigate is how continuous motion affects the image quality of synthetic aperture focusing and to find solutions to reach a fair balance between linear scanning speed and image quality. Additionally, *in-vivo* experiments are not yet conducted to evaluate its performance on human subjects, and therefore effects of human tissue and the motion of human subjects on the performance of the proposed imaging mechanism is still uncharted water.

## VI. Conclusions

In this paper, we proposed a mechanism based on actuated acoustic reflector for 3D US imaging to overcome the limitations in conventional 3D US imaging systems. Unlike a 2D array, the proposed mechanism requires only simple hardware and enjoys lower cost. At the same time, it has a larger FOV compared to a hand-held integrated 1D array and does not require an external actuation unit. The AD-SAF beamforming method is proposed to improve the quality of the 3D US image on the elevation plane. Results from the simulation demonstrate that AD-SAF improves image quality on the elevation plane by achieving lower FWHM and higher SNR. The prototype of the imaging mechanism is built with 38mm (lateral) by 38mm (elevation) by 50mm (axial) FOV and is used to scan fishing-wire phantom and breast phantom. Results from experiments are consistent with simulation studies, providing further validation to the proposed imaging mechanism in its 3D imaging capability and improved image quality on the elevation plane brought by AD-SAF. For future work, we will put emphasis on developing an effective encapsulation solution for the proposed imaging mechanism and improving volume rate to pave the way towards clinical applications.

## Acknowledgments

The authors would like to acknowledge Mr. Enxhi Jaupi from the Department of Biomedical Engineering, Worcester Polytechnic Institute, for his help with manufacturing the metal support frame in the experimental setup, and Mr. Shang Gao from the Department of Robotics Engineering, Worcester Polytechnic Institute, for his help with delay-and-sum and coherent factor implementation.

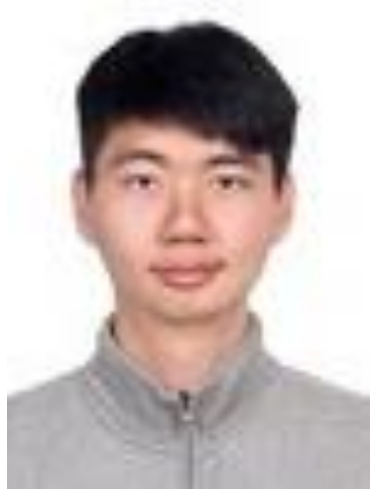

**Yichuan Tang** received BEng with honours in Mechanical Engineering from the University of Nottingham (China campus), in the city of Ningbo in 2016. He received master's degree in Robotics Engineering from Johns Hopkins University in 2018. Since 2019 he has been a PhD student in the department of Robotics Engineering in Worcester Polytechnic Institute, Worcester, Massachusetts, USA. His research focuses on ultrasound and photoacoustic imaging instrumentation. He is affiliated with Medical FUSION Lab directed by Dr. Haichong K. Zhang.

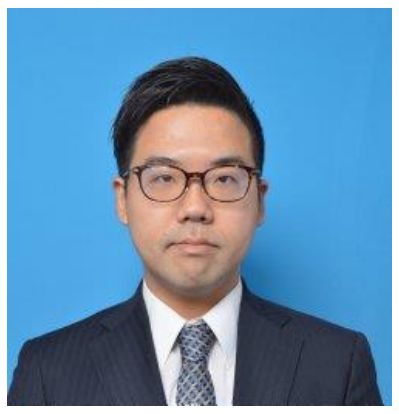

**Ryosuke Tsumura** (M'17) received his BS, MS and PhD in Mechanical Engineering from Waseda University, Tokyo, Japan in 2014, 2016 and 2019, respectively. He is currently a researcher in the Health and Medical Research Institute at the National Institute of Advanced Industrial Science and Technology and an affiliated postdoctoral fellow in Department of Robotics Engineering at the Worcester Polytechnic Institute, MA, USA. His current research interests include robotic intervention and ultrasound imaging.

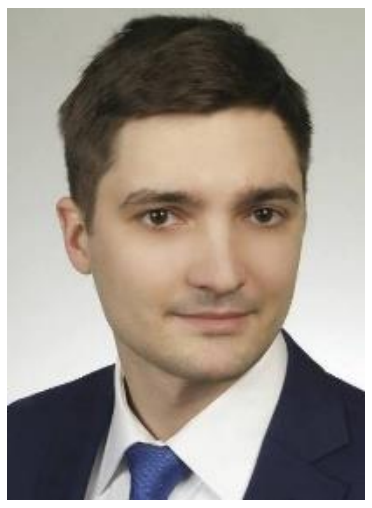

**Jakub Tomasz Kaminski** was born in Poznan, Poland in 1992. He obtained BSc in Mechatronics (2016) and MS in Automatic Control and Robotics (2018) from the Poznan University of Technology, Poland. He is a MS student in the department of Robotics Engineering at Worcester Polytechnic Institute, MA, USA which he joined as a Fulbright Graduate Student Award recipient. Jakub is affiliated with Medical FUSION Lab lead by Dr. Haichong K. Zhang.

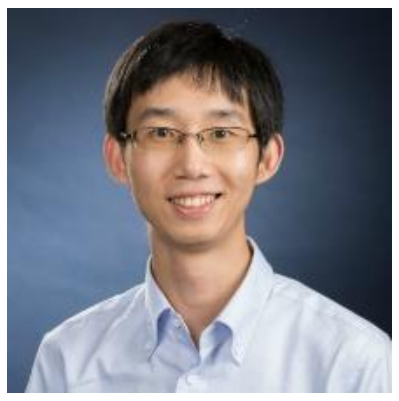

**Haichong (Kai) Zhang** is Haichong (Kai) Zhang is an Assistant Professor in Biomedical Engineering and Robotics Engineering with an appointment in Computer Science at Worcester Polytechnic Institute (WPI). He is the founding director of the Medical Frontier Ultrasound Imaging and Robotic Instrumentation (Medical FUSION) Laboratory. The research in his lab focuses on the interface of medical imaging, sensing, and robotics, developing robotic assisted imaging systems as well as image-guided robotic interventional platforms, where ultrasound and photoacoustic imaging are two key modalities to be investigated and integrated with robotics. Dr. Zhang received his B.S. and M.S. in Human Health Sciences from Kyoto University, Japan, and subsequently earned his M.S. and Ph.D. in Computer Science from the Johns Hopkins University.